\documentclass[aps,prb,twocolumn,showpacs,preprintnumbers,groupedaddress]{revtex4}
\usepackage{graphicx}

\begin{document}

\draft


\title{
One-dimensional continuum and exciton states in quantum wires 
}

\author{
Hidefumi Akiyama${}^{1,2,3}$, Loren N. Pfeiffer${}^2$, Masahiro Yoshita${}^1$, Aron Pinczuk${}^{2,3}$, and Ken W. West${}^2$
}

\affiliation{
${}^1$Institute for Solid State Physics, University of Tokyo, \\
5-1-5 Kashiwanoha, Kashiwa, Chiba 277-8581, Japan\\ 
${}^2$Bell Laboratories, Lucent Technologies, \\
600 Mountain Avenue, Murray Hill, NJ 07974, USA, \\
${}^3$Department of Physics, Columbia University, \\
New York, NY 10027, USA, \\
}

\date{\today}

\begin{abstract}
High-quality T-shaped quantum wires are fabricated by cleaved-edge overgrowth with the molecular beam epitaxy on the interface improved by a growth-interrupt high-temperature anneal. Characterization by micro-photoluminescence (PL) and PL excitation (PLE) spectroscopy at 5 K reveals high uniformity, a sharp spectral width, and a small Stokes shift of one-dimensional (1-D) excitons. The PLE spectrum for 1-D states shows a large peak of ground-state excitons and a small absorption band ascribed to 1-D continuum states with an onset at 11 meV above the exciton peak. 
\end{abstract}

\maketitle

\narrowtext

In semiconductor quantum wires, strong one-dimensional (1-D) Coulomb interactions \cite{Loudon,OgawaPRB91,PKMpp242} cause 1-D excitons to have a large binding energy \cite{WegscheiderPRL,SomeyaPRL,many,SzymanskaPRB} and strong absorption intensity in the ground state \cite{AkiyamaPRB96c}. Moreover, it has been predicted that absorption intensity of continuum states at the band edge should be reduced from electronic joint density of states, which is proportional to inverse square root of energy $1/\sqrt{E}$ in 1-D systems \cite{OgawaPRB91,PKMpp242}. The ratio of the absorption intensity of the continuum states against electronic joint density of states is called the Sommerfeld factor. Thus, the Sommerfeld factor should be less than one in 1-D systems. 

In 2-D and 3-D systems, on the other hand, the Coulomb interaction makes the absorption intensity of the continuum states at the band edges enhanced from the joint density of states \cite{PKMpp242,MillerPRB}. In other words, the Sommerfeld factor is more than one in 2-D and 3-D systems. Such a striking contrast has long been an issue of fundamental interest in low-dimensional structures. 

However, a major problem of practical quantum wires for experimental study of the above effect has been the large energy broadening due to structural inhomogeneities such as interface roughness, which disturbs the subtle detail inherent in 1-D systems. Over the past ten years the cleaved-edge-overgrowth method of molecular beam epitaxy (MBE) has been developed \cite{PfeifferAPL} to fabricate high-quality quantum wires (T-wires) formed at the right angle T-shaped intersection of two quantum wells (QWs) \cite{WegscheiderPRL,SomeyaPRL,GoniAPL,HasenNAT,AkiyamaJP, WegscheiderAPL,RubioSSC}, in which the atomic precision of the quantum wire is determined entirely by the atomic precision of the individual intersecting wells. Moreover we have recently-developed an annealing technique \cite{YoshitaJJAP} that has dramatically improved the (110) interface uniformity. A doped T-wire sample formed with this method has shown an order of magnitude sharper photoluminescence (PL) linewidth compared to previous T-wires \cite{AkiyamaSSC}. 

In this work, we fabricate high-quality non-doped T-wires in a T-wire laser structure, and demonstrate striking improvement in uniformity, spectral width, and Stokes shift by micro-PL, PL excitation (PLE), and scanning micro-PL characterizations. The observed PLE spectrum of the T-wires has shown an isolated strong single peak and a small continuous absorption band with an onset at 11 meV above the peak. Experiments indicate that the peak and the continuous band are intrinsic to quantum wires, so that they are interpreted as the 1-D exciton ground state and 1-D continuum states, or 1-D inonized electron-hole excited states with continuous energy spectrum. In fact, the observed PLE spectrum of T-wires agrees qualitatively with theoretically predicted features inherent to 1-D excitonic systems \cite{ Loudon,OgawaPRB91,PKMpp242}.

\begin{figure}
\includegraphics[width=.3\textwidth]{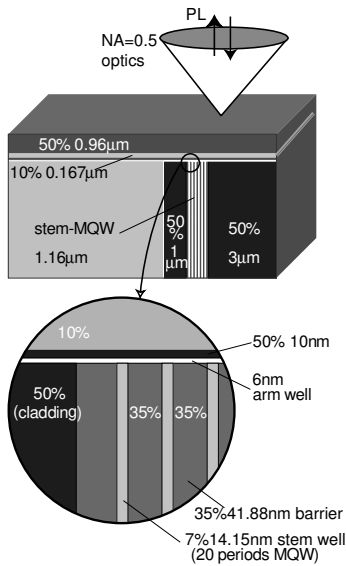}
\caption{
Schematic of a T-wire laser sample. Percentages show Al-concentration $x$ in Al${}_x$Ga${}_{1-x}$As. The laser contains 20 periods of T-wires defined by 7\%-Al filled 14.15nm stem wells and 6nm arm well embedded in a T-shaped optical waveguide with a 500$\mu$m cavity length. Micro-PL and PLE are measured via a 0.5 numerical aperture objective lens through the (110) cleaved-edge overgrowth surface. 
}
\label{1}
\end{figure}

Figure 1 shows a schematic of our T-wire sample, which is embedded in a T-wire laser structure fabricated by the following procedures. 
First, on a non-doped (001) GaAs substrate we successively grew a 50 nm GaAs buffer layer, a 1 $\mu$m cladding layer of 50 \% digital alloy (GaAs)${}_4$(AlAs)${}_4$, 1.16 $\mu$m thick multiple QW layer composed of 20 periods of 14.15 nm Al${}_{0.073}$Ga${}_{0.927}$As QWs (stem well) and 41.88 nm Al${}_{0.35}$Ga${}_{0.65}$As barriers, a 3 $\mu$m cladding layer of 50\% digital alloy (GaAs)${}_4$(AlAs)${}_4$, and a 10 nm GaAs cap layer. The substrate growth temperature was 600~${}^{\circ}$C, and growth was interrupted for 15 seconds after each Al${}_{0.073}$Ga${}_{0.927}$As stem well and each GaAs layer in the 50\% digital alloy.

\begin{figure}
\includegraphics[width=.45\textwidth]{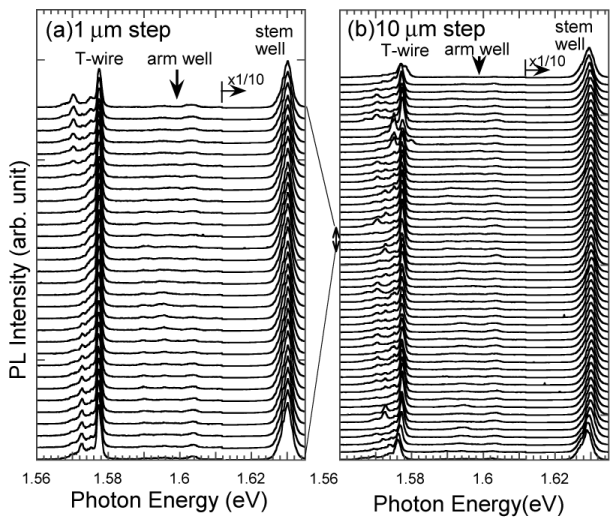}
\caption{
Scanning PL spectra measured in (a) 1$\mu$m steps for 30 $\mu$m and  (b) 10$\mu$m steps for 500 $\mu$m along T-wires at 5K. Spot size was about 0.8 $\mu$m in diameter. The measured 30 $\mu$m region of (a) is indicated by side arrows in (b). 
}
\label{2}
\end{figure}

Then, in a separate MBE growth on an {\it in-situ} cleaved (110) edge of this structure, we grew at 490~${}^{\circ}$C by the cleaved-edge overgrowth method, a 6 nm GaAs QW (arm well) layer, a 10 nm Al${}_{0.5}$Ga${}_{0.5}$As barrier layer, a 167nm Al${}_{0.1}$Ga${}_{0.9}$As core layer, a 0.96 $\mu$m cladding layer of 50 \% digital alloy (GaAs)${}_6$(AlAs)${}_6$, and a 10 nm GaAs cap layer. Right after the growth of the GaAs arm well layer, growth was interrupted for a 10 minute anneal at 600~${}^{\circ}$C. This anneal is based on our recently-developed growth-interruption annealing technique \cite{YoshitaJJAP}. 

Substrate rotation was performed by 10 revolutions per minute during the MBE growths both in the (001) and the (110) directions. Our fabrication method is similar to that in our previous work \cite{WegscheiderPRL,RubioSSC} except for the anneal step, that dramatically reduces interface roughness on (110)-GaAs surfaces. 

Twenty periods of highly uniform T-wires are thus formed at the T-shaped intersections of 7\%-Al-filled 14nm stem wells and a 6 nm arm well, and embedded in the core of T-shaped optical waveguide. A laser bar with 500$\mu$m cavity length was cut from the wafer by cleavage, attached on a copper block with silver paint, and cooled down to 5K on the cold finger of a helium-flow-type cryostat. Lasing properties of the sample will be reported elsewhere \cite{AkiyamaPRL2}.

Micro-PL and PLE measurements on the T-wires were performed with a cw titanium-sapphire laser via the overgrowth (110) surface in the geometry of point excitation into a 0.8 $\mu$m spot \cite{YoshitaJAP} centered at the T-wire region with a 0.5 numerical aperture objective lens. A 0.6m triple spectrometer with a 600 grooves/mm grating and a back-illumination-type liquid-nitrogen-cooled charge-coupled-device camera were used to detect PL. Spectral resolution was 0.3 meV.

Figure 2 (a) shows scanning micro-PL spectra by 1 $\mu$m steps over 30 $\mu$m along T-wires at 5 K. The excitation light had intensity of 2 $\mu$W and photon energy of 1.653 eV. Well-resolved sharp peaks are observed for the T-wires (1.578 eV) and the stem well (1.630 eV). PL of arm well is not observed, because carriers created in the arm well quickly flow into T-wires. The full-width-of-half-maximum (FWHM) of the T-wire peak is 1.5 meV, far smaller than those reported in previous work \cite{WegscheiderPRL,SomeyaPRL,AkiyamaPRB96c,GoniAPL,HasenNAT,AkiyamaJP,WegscheiderAPL,RubioSSC}, showing the high flatness of the interfaces. 

PL peak of stem wells and the main peak of T-wires are very uniform over 30 $\mu$m. Intensities of small peaks in the lower energy side of the main peak of T-wires vary from place to place. Thus, these peaks are ascribed to localized excitons in T-wires. The PL energy positions of the localized excitons are discrete with separation of 2.4meV. This is consistent with thickness fluctuation of arm well by integer monolayers. Atomic force microscope study of arm well surfaces has shown the existence of islands with specific shape and 1-3 monolayer heights \cite{YoshitaJJAP}. Small intensities of the localized exciton peaks indicate that such islands are very rare. 

Figure 2 (b) shows scanning micro-PL spectra by 10 $\mu$m steps over 500 $\mu$m along T-wires at 5 K, which is the whole region of the T-wire laser cavity. The 30 $\mu$m region shown in Fig. 2 (a) is represented by the central three spectra indicated by the side arrows. The whole spectra show that PL of T-wires is uniform over 500 $\mu$m. 

\begin{figure}
\includegraphics[width=.35\textwidth]{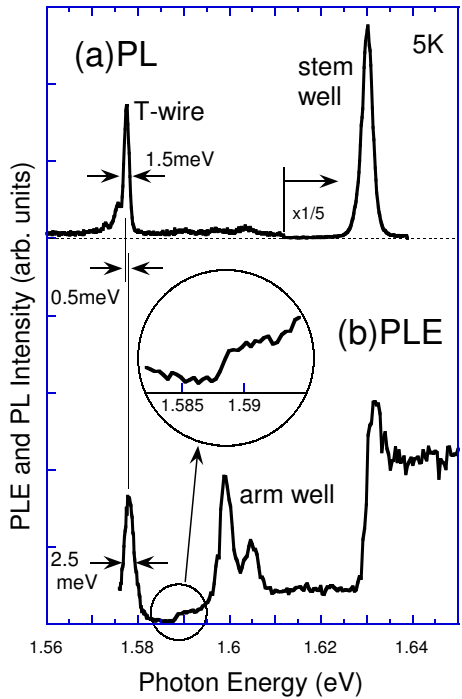}
\caption{
PL (a) and PLE (b) spectra of T-wires at 5K measured via point spectroscopy with a spot size of about 0.8 $\mu$m diameter. The PL and PLE widths of T-wires are 1.5meV and 2.5meV, respectively, and the Stokes shift is 0.5meV. The inset is a blowup of the PLE spectrum around the continuous absorption band starting at 1.589 eV. It is ascribed to the excited levels of 1-D excitons, namely the second and higher bound states of 1-D excitons and the 1-D continuum states in T-wires. 
}
\label{3}
\end{figure}

\begin{figure}
\includegraphics[width=.4\textwidth]{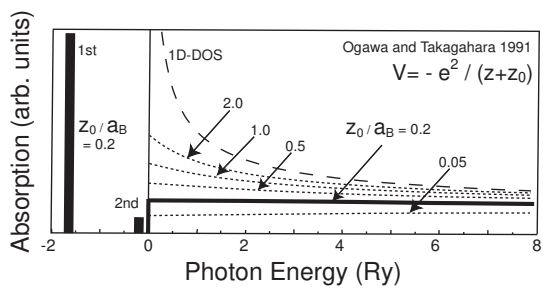}
\caption{
Characteristics of absorption spectra of 1-D excitons and 1-D continuum reproduced from calculated results by Ogawa and Takagahara \cite{OgawaPRB91} for the Coulomb potential $V=-e^2/(z+z_0)$,  where $z$, $z_0$, $a_B$, and $R_y$ show electron-hole distance, its cutoff length, 3-D Bohr radius, and 3-D Rydberg energy for excitons, respectively.
A dashed curve shows 1-D density of states $1/\sqrt{E}$. The strong 1-D Coulomb interaction moves the oscillator strength out of the low energy edge of the continuum states into the stabilized ground-state exciton. 
}
\label{4}
\end{figure}

Figure 3 (a) shows PL spectrum at 5 K for a spot at the central region of Fig. 2 (a). Figure 3 (b) shows PLE spectrum of the T-wire PL, which reveals the absorption spectrum of the T-wires. The excitonic absorption peaks of T-wire (1.578 eV), arm well (1.599 eV), and stem well (1.632 eV) are clearly observed. The two peaks at 1.599 eV and at 1.605 eV are due to the arm well in the core region on 35\% barrier and the adjacent arm well in the cladding region on 50\% barrier, respectively, which is confirmed by a separate PL imaging experiment. 

The Stokes shift of ground state excitons in T-wires, that is the peak shift energy of PL and PLE peaks for T-wires, is found to be 0.5meV, whereas the FWHM width of T-wire PLE peak is 2.5meV. The Stokes shift is small compared with the width, so that the PL peak is well overlapped with the PLE peak. This demonstrates that the PL of T-wires has a free-exciton nature. This is the first such demonstration for quantum wires. 

Note that a continuous absorption band is observed in the PLE spectrum with an onset at 1.589 eV, as magnified in the inset of Fig. 3. It is 11 meV above the PL peak of the ground-state free exciton in T-wires. Such an absorption band was observed in all of several different pieces of samples that we measured. In particular, a single quantum wire sample with only one wire instead of 20 wires showed a continuous absorption band with a similar shape but with significantly smaller intensity than the present 20 wire sample. The reproducibility and the correlation of its intensity with wire numbers suggest that the continuous absorption band is not caused by extrinsic low energy states in the arm well but originates from intrinsic states in quantum wires. 

In general, an exciton peak should accompany a continuous absorption band separated by an exciton binding energy, because excitons not only have a bound ground state but also have ionized continuous excited states of an electron and a hole. Thus, a continuous absorption band should be observed together with an exciton peak in a PLE measurement with high resolution and sensitivity. Therefore, the observed continuous absorption band in Fig. 3 is expected to correspond to ionized continuous excited states of an electron and a hole in T-wires, namely 1-D continuum states. 
A similar measurement of the excited levels of 2-D excitons in quantum wells was reported in 1981 by Miller and coworkers \cite{MillerPRB}, in which they resolved the 1s-like ground-state exciton and the 2-D continuum states with a line shape close to the step-function-like 2-D density of state. Our work is the first such observation of 1-D continuum states in quantum wires enabled by improvement in spectral sharpness of the wires. 

It is remarkable that the 1-D continuum states do not appear to have a $1/\sqrt{E}$-like singularity at the onset as would be expected from a 1-D joint density of states. Moreover, the peak intensity of the 1-D exciton ground state is much stronger than the intensity of the onset of the 1-D continuum states in quantum wires. 
In fact, it has been theoretically predicted \cite{Loudon,OgawaPRB91,PKMpp242} for 1-D excitonic systems that the exciton ground state has strong absorption intensity in the ground state. Instead, absorption intensity of continuum states at the band edge should be reduced, and singularity in 1-D electronic joint density of states proportional to $1/\sqrt{E}$ should be removed.

To discuss this predicted effect, we made, in Fig. 4, combined plots of absorption spectra of 1-D excitons and continuum states calculated by Ogawa and Takagahara in 1991 \cite{OgawaPRB91}. Neglecting real confinement potential, they solved an 1-D schr\"odinger equation for the Coulomb potential $-e^2/(z+z_0)$, where $z$ and $z_0$ are the electron-hole distance and an artificial cutoff length. Two thick vertical lines in the region of negative photon energy represent positions and relative intensity for absorption peaks of the ground and second exciton bound states for $z_0/a_B=0.2$, where $a_B$ shows the 3-D exciton Bohr radius. A thick curve in the region of positive photon energy represents absorption profile due to 1-D continuum states for $z_0/a_B=0.2$. In comparison, 1-D density of states proportional to $1/\sqrt{E}$ are shown by a dashed curve, which reveals that the absorption intensity of the 1-D continuum states is suppressed at the band edge. For smaller choice of $z_0$, the ground-state-exciton peak become stronger in intensity and lower in energy (not shown), while 1-D continuum states become smaller as shown by dotted curves for $z_0/a_B$=2.0, 1.0, 0.5, and 0.05. The second exciton peak is not sensitive to $z_0$. The physics suggested by this theory is that strong 1-D Coulomb interactions move the oscillator strength out of the low energy edge of the continuum states into the stabilized ground-state exciton\cite{PKMpp242}. Thus, the $1/\sqrt{E}$ singularity of the 1-D free-particle density of states is completely suppressed in the line shape for 1-D continuum states, and the shape becomes very similar to a step function. 

Our result in Fig. 3 is the first experimental data in semiconductor quantum wires to be compared with such theoretical predictions \cite{Loudon,OgawaPRB91,PKMpp242}, where the 1-D ground state exciton peak and a continuum absorption band are completely resolved. 

For more quantitative discussion, we need to compare our result with more practical calculations. Though there are many theoretical papers\cite{many} on 1-D exciton binding energies in T-wires, calculation of absorption spectra as well as binding energies to be compared with our present experiment has been reported only by Szymanska and coworkers \cite{SzymanskaPRB,RubioSSC}. 
The calculated spectrum \cite{SzymanskaPRB,RubioSSC} shows a strong peak of the 1-D exciton ground state, a weaker peak of the second exciton state, and a series of continuously populated small states. The continuous states have onset at 11 meV above the ground states and have fairly flat intensity, which are in good agreement with our experimental results. On the other hand, the peak of the second exciton states appeared in the calculated spectrum at 6 meV above the ground state is not observed in our experiment. It suggests that further work both in measurements and theoretical calculations is necessary. 

In summary, we achieved dramatic improvement in uniformity, spectral width, and Stokes shift in T-shaped quantum wires by growth interrupt anneal with cleaved edge overgrowth in MBE. The observed PLE spectrum of the T-wires shows the strong ground-state-exciton peak and a small continuous absorption band with an onset at 11 meV above the peak, which is ascribed to 1-D continuum states. The observed PLE spectrum of T-wires agrees qualitatively with theoretically predicted features inherent to 1-D excitonic systems meaning that the $1/\sqrt{E}$ singularity of the 1-D free-particle density of states is suppressed in the line shape for 1-D continuum states as a result of the strong 1-D Coulomb interaction effect. 

We thank Professor P. B. Littlewood and Dr. M. Szymanska in University of Cambridge for valuable discussion, and Prof. T. Ogawa in Osaka University for discussion and permission for using his published results. One of the authors (H. A.) acknowledges the financial support from the Ministry of Education, Culture, Sports, Science and Technology, Japan.


\begin{references}


\bibitem{Loudon}
R. Loudon, Am. J. Phys. {\bf 27}, 649 (1959); R. J. Eliott and R. Loudon, J. Phys. Chem. Solids {\bf 15}, 196 (1960). 

\bibitem{OgawaPRB91} T. Ogawa and T. Takagahara, Phys. Rev. B {\bf 43}, 14325 (1991); {\bf 44}, 8138 (1991). 

\bibitem{PKMpp242} N. Peyghambarian, S. W. Koch, and A. Mysyrowicz, {\it Introduction to Semiconductor Optics}, (Prentice-Hall Inc., New Jersey, 1993) pp. 150, 217, and 242. 

\bibitem{WegscheiderPRL} W. Wegscheider, L. N. Pfeiffer, M. M. Dignam, A. Pinczuk, K. W. West, S. L. McCall, and R. Hull, Phys. Rev. Lett. {\bf 71}, 4071 (1993).

\bibitem{SomeyaPRL} T. Someya, H. Akiyama, and H. Sakaki, Phys. Rev. Lett. {\bf 74}, 3664 (1995); {\bf 76}, 2965 (1996).

\bibitem{many} 
F. Rossi, G. Goldoni, and E. Molinari, Phys. Rev. Lett. {\bf 78}, 3527 (1997); {\bf 80}, 4995 (1998). 
A. Thilagam, J. Appl. Phys. {\bf 82}, 5753 (1997). 
S. Glutsch, F. Bechstedt, W. Wegscheider, and G. Schedelbeck, Phys. Rev. B {\bf 56}, 4108 (1997). 
D. Brinkmann and G. Fishman, {\it ibid.} {\bf 56}, 15 211 (1997). 
S. N. Walck, T. L. Reinecke, and P. A. Knipp, {\it ibid.} {\bf 56}, 9235 (1997). 
Y. Zhang and A. Mascarenhas, {\it ibid.} {\bf 59}, 2040 (1999). 
M. Stopa, {\it ibid.} {\bf 63}, 195312 (2001). 

\bibitem{SzymanskaPRB}
M. H. Szymanska, P. B. Littlewood, and R. J. Needs, Phys. Rev. B {\bf 63}, 205317 (2001). 

\bibitem{AkiyamaPRB96c}
H. Akiyama, T. Someya, and H. Sakaki, Phys. Rev. B {\bf 53}, R16160 (1996)

\bibitem{MillerPRB}
R. C. Miller, D. A. Kleinman, W. T. Tsang, and A. C. Gossard, Phys. Rev. B {\bf 24}, 1134 (1981). 

\bibitem{PfeifferAPL} L. Pfeiffer, K. W. West, H. L. St\"ormer, J. P. Eisenstein, K. W. Baldwin, D. Gershoni, and J. Spector, Appl. Phys. Lett. {\bf 56}, 1697 (1990).

\bibitem{GoniAPL} A. R. G\~oni, L. N. Pfeiffer, K. W. West, A. Pinczuk, H. U. Baranger, and H. L. St\"ormer, Appl. Phys. Lett. {\bf 61}, 1956 (1992).

\bibitem{HasenNAT} J. Hasen, L. N. Pfeiffer, A. Pinczuk, S. He, K. W. West, and B. S. Dennis, Nature {\bf 390}, 54 (1997).

\bibitem{AkiyamaJP} H. Akiyama, J. Phys.: Condens. Matter {\bf 10}, 3095 (1998).

\bibitem{WegscheiderAPL} W. Wegscheider, L. Pfeiffer, K. West, and R. E. Leibenguth, Appl. Phys. Lett. {\bf 65}, 2510 (1994).

\bibitem{RubioSSC} J. Rubio, L. Pfeiffer, M. H. Szymanska, A. Pinczuk, Song He, H. U. Baranger, P. B. Littlewood, K. W. West, and B. S. Dennis, Solid State Commun. {\bf 120}, 423 (2001). 

\bibitem{YoshitaJJAP} M. Yoshita, H. Akiyama, L. N. Pfeiffer, and K. W. West, Jpn. J. Appl. Phys. {\bf 40}, L252 (2001); Appl. Phys. Lett. {\bf 81}, 49 (2002). 

\bibitem{AkiyamaSSC}H. Akiyama, L. N. Pfeiffer, A. Pinczuk, K. W. West, and M. Yoshita, Solid State Commun. {\bf 122}, 169 (2002). 

\bibitem{AkiyamaPRL2}H. Akiyama, L. N. Pfeiffer, M. Yoshita, A. Pinczuk, K. W. West, M. J. Matthews, and J. Wynn, unpublished. 

\bibitem{YoshitaJAP} M. Yoshita, H. Akiyama, T. Someya, and H. Sakaki, J. Appl. Phys. {\bf 83}, 3777 (1998). 




\end{references}
\end{document}